\documentclass[aps,prb,preprint,superscriptaddress,showpacs]{revtex4-1}

\usepackage{graphicx}% Include figure files
\usepackage{dcolumn}% Align table columns on decimal point
\usepackage{bm}% bold math
\usepackage{amsmath}
\usepackage{ulem}
\usepackage[caption=false]{subfig}
\usepackage{color,soul}

\begin{document}

%\preprint{}

\title{Impact of further-range exchange and cubic anisotropy on magnetic excitations in the fcc kagome antiferromagnet IrMn$_3$}


%\author{}
%\thanks{}
%\affiliation{}
\author{M. D. LeBlanc}
%\thanks{}
\affiliation{Department of Physics and Physical Oceanography, Memorial University of Newfoundland, St. John's, Newfoundland, A1B 3X7,  Canada}

\author{A. A. Aczel}
%\thanks{}
\affiliation{Oak Ridge Natl Lab, Neutron Scattering Div, Oak Ridge, TN 37831 USA}

\author{G. E. Granroth}
%\thanks{}
\affiliation{Oak Ridge Natl Lab, Neutron Scattering Div, Oak Ridge, TN 37831 USA}

\author{B. W. Southern}
%\thanks{}
\affiliation{Department of Physics and Astronomy, University of Manitoba, Winnipeg, MB, R3T 2N2, Canada}

\author{J.-Q. Yan}
%\thanks{}
\affiliation{Oak Ridge Natl Lab, Div Mat Sci $\&$ Technol, Oak Ridge, TN 37831 USA}

\author{S. E. Nagler}
%\thanks{}
\affiliation{Oak Ridge Natl Lab, Neutron Scattering Div, Oak Ridge, TN 37831 USA}

\author{J. P.	 Whitehead}
%\thanks{}
\affiliation{Department of Physics and Physical Oceanography, Memorial University of Newfoundland, St. John's, Newfoundland, A1B 3X7,  Canada}

\author{M. L. Plumer}
%\thanks{}
\affiliation{Department of Physics and Physical Oceanography, Memorial University of Newfoundland, St. John's, Newfoundland, A1B 3X7,  Canada}

%\author{THIS IS MP's  PROPOSED ORDERING FOR AUTHORSHIP AND STILL NEEDS TO BE AGREED UPON}
%\thanks{}
%\affiliation{Department of Physics and Physical Oceanography, Memorial University of Newfoundland, St. John's, Newfoundland, A1B 3X7,  Canada}
%\author{J. P. Whitehead}
%\thanks{}
%\affiliation{Department of Physics and Physical Oceanography, Memorial University of Newfoundland, St. John's, Newfoundland, A1B 3X7,  Canada}


\date{\today}

\begin{abstract}
Exchange interactions up to fourth nearest neighbor are shown within a classical local-moment Heisenberg approach to be important to model inelastic neutron scattering data on the fcc kagome antiferromagnet IrMn$_3$.  Spin wave frequencies are calculated using the torque equation and the magnetic scattering function, $S({\bf Q},\omega)$, is determined by a Green's function method, as an extension of our previous work, LeBlanc {\it et al.} Phys. Rev. B {\bf 90}, 144403 (2014).   
Results are compared with intensity contour data on powder samples of ordered IrMn$_3$, where magnetic Mn ions occupy lattice sites of ABC stacked kagome planes. Values of exchange parameters taken from DFT calculations used in our model provide good agreement with the experimental results only if further-neighbor exchange is included.  Estimates of the observed energy gap support the existence of strong cubic anisotropy predicted by DFT calculations.
  
\end{abstract}

%\keywords{}

\maketitle

\section{INTRODUCTION}

IrMn$_3$ provides an important example of a truly three dimensional (3D) kagome lattice giving rise to geometrical magnetic frustration from eight near-neighbor (NN) antiferromagnetic exchange interactions between Mn ions.\cite{hemmati2012} ABC stacking of kagome planes of Mn ions in the cubic $\langle 111 \rangle$ directions gives an overall L1$_2$, AuCu$_3$-type, fcc structure with four NNs within each plane and two NN connecting each adjacent plane (see Fig. 1).
Interest in the magnetic properties of the corresponding two-dimensional (2D) kagome NN Heisenberg antiferromagnet spans 25 years due to the macroscopic spin degeneracy of the basic 120$^\circ$ spin structure associated with corner-sharing triangles.\cite{chalker1992} The zero energy dispersionless (flat) spin wave mode at all wave vectors predicted from classical theory acquires dispersion in the presence of further-neighbor exchange interactions.\cite{harris1992,schnabel2012} Inelastic neutron scattering data on a system with weakly coupled kagome planes appear consistent with this scenario where the flat mode observed is gapped due to additional Dzyaloshinskii-Moriya interactions.\cite{matan06} 

\begin{figure}[htp!]
\centering
\includegraphics[width=0.55\textwidth]{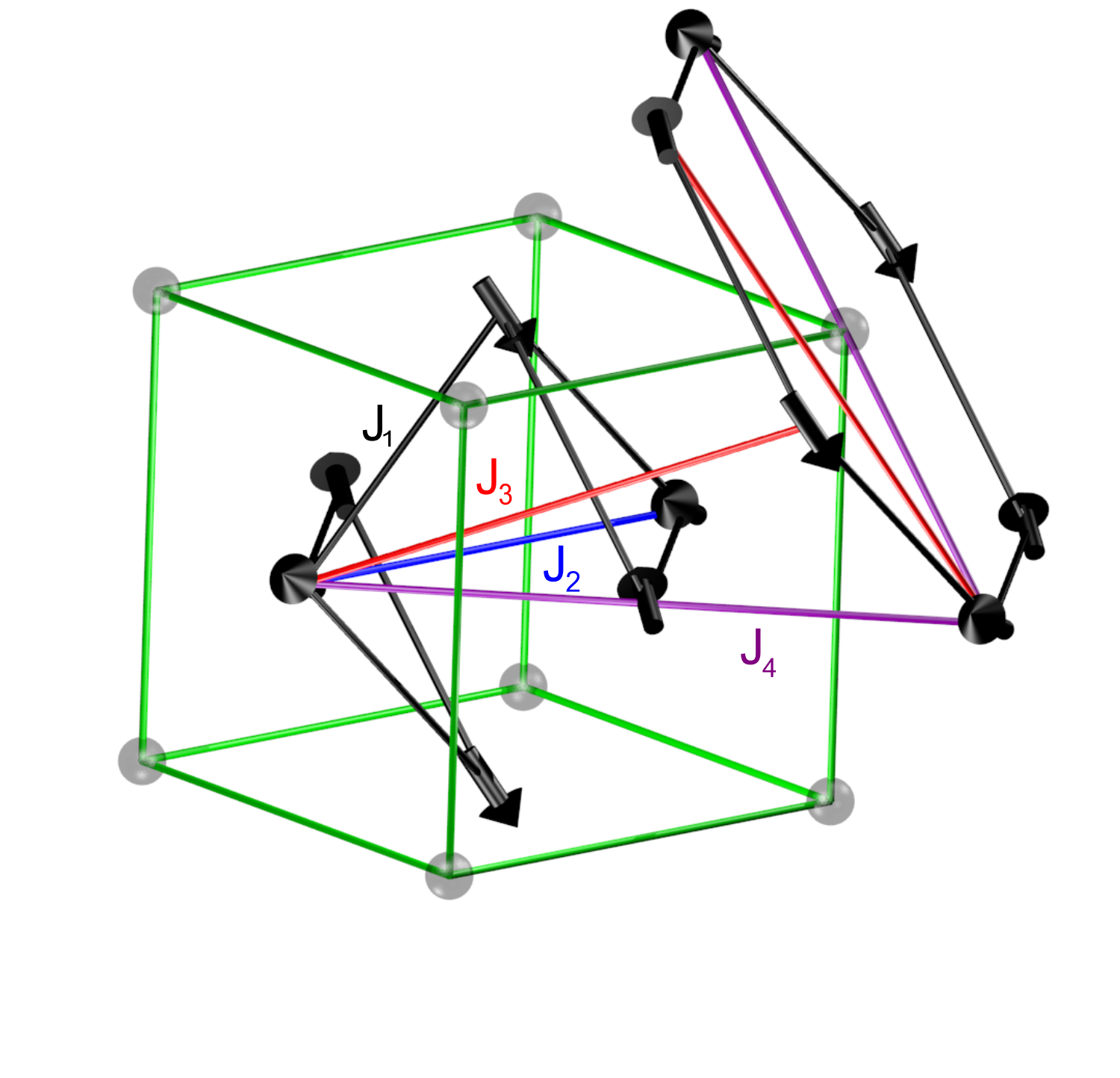}
\caption{ABC stacked kagome planes forming the fcc kagome lattice with magnetic ions occupying the cube face center sites. Illustrated are spin vectors forming the 120$^\circ$ q=0 [111] planar spin structure (zero anisotropy) and the 
exchange interactions J$_1$, J$_2$, J$_3$, and J$_4$.  Also see Table I.}\label{Fig1}
\end{figure}

Interest in IrMn$_x$ alloys over the past 15 years has mainly arisen due to applications in spin-valve technology, where they have been widely used as the antiferromagnetic thin film of choice that pins the magnetic moments of an adjacent ferromagnet in the phenomenon known as exchange bias.\cite{ogrady2010,tsunoda10,kohn2013,yanes2013} Although there is no universally accepted microscopic mechanism for exchange bias, magnetic frustration is believed to be important and stoichiometric IrMn$_3$ appears to optimize the desired pinning.\cite{tsunoda10}

Monte Carlo simulations of the NN Heisenberg fcc kagome lattice have shown that the basic co-planar 120$^\circ$ q=0 magnetic structure observed in 2D persists in the 3D case, with the inter-spin angle being 120$^\circ$ between all eight NNs (shown in Fig. 1).\cite{hemmati2012}  The spin degeneracy is reduced in 3D and exists in the form of sublattice magnetization switching in the stacked kagome planes.  This persistent degeneracy is believed to be responsible for the first-order nature of the phase transition at $T_N$.  Inspired by earlier electronic structure calculations,\cite{szunyogh2009} subsequent simulations of the 3D lattice with an effective local cubic anisotropy included provided evidence that anisotropy removes the basic kagome degeneracy and the structure becomes non co-planar with a finite magnetization (spin vectors are lifted out of the [111] plane).  This release of frustration drives the transition to be continuous.\cite{leblanc2013}  This scenario was supported by spin wave calculations of the NN Heisenberg model with and without anisotropy which demonstrated that in the absence of anisotropy, the zero energy flat mode exists only in certain high symmetry directions in reciprocal space and that the addition of anisotropy induces a gap (Ref.\onlinecite{leblanc2014}, hereafter referred to as I).  Monte Carlo studies of [111] thin films confirmed that the q=0 magnetic order remains in these geometries and the impact of surface axial anisotropy was also considered.\cite{yerzhakov2016} 

Early neutron scattering experiments on sister compounds RhMn$_3$ and PtMn$_3$ revealed the 120$^\circ$ magnetic order\cite{kren66} which was subsequently established also in single crystal neutron scattering studies on IrMn$_3$ by Tomeno et al.\cite{tomeno1999} and referred to as ``T1'' magnetic order, with no mention of the underlying kagome lattice structure of magnetic ions or any indication of a finite magnetization. More recently, single crystal [111] thin films of IrMn$_3$ also showed the same magnetic order as in the bulk where exchange bias was also studied.\cite{kohn2013} To our knowledge, there have been no  neutron scattering experiments reporting on spin excitations in these magnetic fcc kagome systems.  It is of interest to note another class of Mn-based compounds with the generic formula Mn$_3$AX also exhibits fcc kagome magnetism.\cite{kaneko1987}

\begin{table}[t!]
\begin{center}
\begin{tabular}{| c | c | c | c | c|}
\hline
\multicolumn{5}{|c|}{ {Exchange Interactions in IrMn$_3$}} \\\hline
nth Near Neighbor & J$_1$  & J$_2$  &  J$_3$ & J$_4$  \\\hline
Value from DFT (meV)     & 40 & -5 & 10 & -5 \\\hline
Neighbors in plane    & 4 & 0 & 4  & 6 \\\hline
Neighbors first adjacent planes   & 4 & 6 & 8 & 0 \\\hline
Neighbors second adjacent planes   & 0 & 0 & 4 & 6 \\\hline
Total     & 8 & 6 & 16 & 12 \\\hline
Vector   & [$\frac{1}{2}$ $\frac{1}{2}$ 0] & [1 0 0] & [1 $\frac{1}{2}$ $\frac{1}{2}$] & [1 1 0] \\\hline
Distance (a)   & 0.707 & 1 & 1.225 & 1.414 \\\hline
\end{tabular}
\vspace{0.5cm}
\caption{Exchange parameters for the fcc kagome lattice IrMn$_3$ (see Fig. 1) where $J > 0$ implies antiferromagnetic coupling.  DFT values are taken from Ref. [\onlinecite{szunyogh2009}].  Distances are relative to the lattice constant\cite{tomeno1999} a=3.76 \AA.}\label{tab:Parameters}
\end{center}
\end{table}

The high N\'eel temperature in IrMn$_3$, $T_N \simeq$ 960 K (with large values also reported in the sister compounds), attractive for device applications, can be associated with large exchange interactions.  For example, NN $J_1 \simeq$ 40 meV (here, J $>$ 0 implies an antiferromagnetic interaction) has been estimated for IrMn$_3$ from the density functional theory (DFT) calculations by Szunyogh et al.\cite{szunyogh2009}, where an effective spin S=2.2 has been folded into the reported exchange constant values.\cite{szunyogh2009,szunyogh2017}  This and related work also reported substantial longer range exchange interactions of an oscillatory nature, with  $J_2 \simeq$ -5 meV, $J_3 \simeq$ 10 meV, $J_4 \simeq$ -5 meV, as well as a large effective cubic anisotropy \cite{szunyogh2011} $K_{\mathrm{eff}}$ $\simeq$ 7.67 meV (see Table I). The DFT results indicate that further  exchange interactions beyond fourth-neighbor are negligbly small.\cite{szunyogh2009} Analysis shows that these longer range alternating antiferromagnetic and ferromagnetic exchange interactions are consistent with the 3D $q=0$ spin structure and do not introduce additional frustration.  A first-principles molecular spin dynamics study of PtMn$_3$ and RhMn$_3$ also reports enhanced second-neighbor exchange interactions.\cite{uchida2016}

In the present work, the earlier local-moment spin-wave and inelastic magnetic scattering intensity calculations reported in I are expanded to include the further-neighbor exchange $J_2$, $J_3$, and $J_4$.  As in the 2D case, these additional interactions remove any flat modes and dispersion appears in all cases.  The impact of cubic anisotropy, K, is again examined in the presence of the additional exchange interactions. Scattering intensity, $S({\bf Q},\omega)$, contours are calculated for both single crystal and powder sample scenarios.  Results are compared with new inelastic neutron scattering data on powder samples of ordered phase IrMn$_3$. The effects of further-order exchange are demonstrated to be important and an energy gap is observed.

\section{MODEL RESULTS}

Modifications to our previous analysis in I to include further-neighbor exchange interactions are described here.  In that work,  J  represented the NN exchange in kagome planes and  J$'$ denoted inter-plane exchange.  Here, we set J$_1$ = J = J$'$.
The 120$^\circ$ q=0 spin structure is characterized by a three magnetic sublattice magnetization vectors labeled as A, B and C, and we consider the following Hamiltonian
\begin{equation}
 {\cal H} = \sum_{i<j} J(|{\bf r}_i- {\bf r}_j|)  \mathbf{S}_i \cdot \mathbf{S}_j - K\sum_{\gamma}\sum_{l\subset \gamma} (\mathbf{S}_l \cdot \mathbf{n}_{\gamma})^2.  
\end{equation}
Note that\cite{leblanc2013}  $K = \frac{1}{2}K_{\mathrm{eff}} \ge 0$ has a different easy direction for each of the three sublattices. Here,  $\gamma$ represents sublattice A, B and C and $l$ is summed over the $\frac{N}{3}$ spins of sublattice $\gamma$, $\mathbf{S}_i$ are unit classical Heisenberg spin vectors at each site and  $\mathbf{n}_\gamma$ are unit vectors in the cube axes directions, $\mathbf{n}_{\rm A}=\hat{\bf{x}}$, $\mathbf{n}_{\rm B}=\hat{\bf{y}},$ and $\mathbf{n}_{\rm C}=\hat{\bf{z}}$.  The lattice constant of ordered IrMn$_3$ has been determined at room temperature to be around a=3.76 \AA.\cite{tomeno1999}  The Mn ions occupy face center sites, with $\gamma$ = A at [0,$\frac{1}{2},\frac{1}{2}$], $\gamma$ = B at [$\frac{1}{2},0,\frac{1}{2}$], and $\gamma$ = C at [$\frac{1}{2},\frac{1}{2}$, 0], separated by a distance a/$\sqrt{2}$ = 2.67 \AA ~ as depicted in Fig. 1.
The model results presented below include the exchange constants up to fourth near-neighbors, with values given in Table I, as well as anisotropy.  In order to demonstrate the impact of these further-neighbor exchange interactions, as well as anisotropy, we also consider model results with some of the constants set to zero, or anisotropy set to zero. We do not attempt to fit model parameters with the data described below. 

As described in I, in the absence of anisotropy spins are coplanar with zero net magnetization.  Anisotropy serves to lift the spin vectors out of the plane and induce a finite magnetization in a $\langle 111\rangle$ direction.\cite{leblanc2013} This effect is characterized by $\alpha$, the cosine of the angle between each sublattice spin and its anisotropy axis ($\alpha = \cos(\boldsymbol{S}_i \cdot {\bf n}_i), i= A,B,C$), and $\beta$, the cosine of the angle with respect to the other two anisotropy axes ($\beta = \cos(\boldsymbol{S}_i \cdot{\bf n}_j)$, $i \neq j$) where $\alpha^2 + 2 \beta^2 =1$. 
The modified ground state energy per spin that includes further-neighbor exchange is given by (compare Eqs. 2 and 3 in I where S=1 was used for convenience)
\begin{equation}
E/N = 4(J_1 + 2J_3)(\beta^2-2\alpha\beta)-K\alpha^2 + 3J_2 + 6J_4
\end{equation}
and is minimized when $\alpha$ has the value 
\begin{equation}
 \alpha = \sqrt{1/2 + 1/2\sqrt{1-1/[1+(\tilde{K}+1)^2/8]}}
\end{equation} where $\tilde{K}= K/(2J_1 + 4J_3)$
and $ \beta = \sqrt{\frac{1-\alpha^2}{2}} $
using the positive values of the square roots to give physical solutions. Note that there are eight degenerate ground states corresponding to the eight $\langle 111\rangle$ axes.\cite{leblanc2013}
The analysis of spin excitations in this section correspond to a single domain [111] crystal. Powder averaged results are discussed in the following section.

The basic structure of the linearized spin wave theory presented in I remains the same with further-neighbor exchange added.  The 6 $\times$ 6 matrix characterizing dynamic fluctuations of the transverse spin components in a local coordinate system is again given by
\begin{equation}
-i\omega \left( \begin{array}{c} {\tilde{S}}_A^x \\ {\tilde{S}}_B^x \\ {\tilde{S}}_C^x \\ {\tilde{S}}_A^y \\ {\tilde{S}}_B^y \\ {\tilde{S}}_C^y \end{array}\right)  =  \left( \begin{array}{cccccc} 0 & Y_{AB} & -Y_{AC} & X & Z_{AB} &Z_{AC} \\
                                                             -Y_{AB} & 0 & Y_{BC} & Z_{AB} & X & Z_{BC} \\
                                                             Y_{AC} & - Y_{BC} & 0 & Z_{AC} & Z_{BC} & X \\
                                                              W & T_{AB} & T_{AC} & 0 & Y_{AB} & -Y_{AC} \\
                                                              T_{AB} & W & T_{BC} & -Y_{AB} & 0 & Y_{BC} \\
                                                              T_{AC} & T_{BC} & W & Y_{AC} & -Y_{BC} & 0 \end{array}\right)
\left( \begin{array}{c} \tilde{S}_A^x \\ \tilde{S}_B^x \\ \tilde{S}_C^x \\ \tilde{S}_A^y \\ \tilde{S}_B^y \\ \tilde{S}_C^y \end{array}\right) 
\end{equation}
where Y, T, and Z are defined in I and with the following modifications:
\begin{equation}
\begin{split}
X =& [8(J_1+ 2J_3)(\beta-2\alpha)\beta -2K\alpha^2  \\
  + & 6J_2 - 2J_2(\cos Q_xa + \cos Q_ya + \cos Q_za) \\
  + & 12J_4 - 4J_4(\cos Q_xa \cos Q_ya +  \cos Q_xa \cos Q_za  + \cos Q_ya \cos Q_za)]/S\\
W =& [8(J_1 + 2J_3)(2\alpha-\beta)\beta +2K(\alpha^2-2\beta^2) \\
  - & 6J_2 + 2J_2(\cos Q_xa + \cos Q_ya + \cos Q_za) \\
  - & 12J_4 + 4J_4(\cos Q_xa \cos Q_ya +  \cos Q_xa \cos Q_za  + \cos Q_ya \cos Q_za)]/S\\
\end{split}
\end{equation}
and
\begin{equation}
\begin{split}
\lambda_{AB} =& 4S^{-1}[J_1 + 2J_3 \cos Q_za] \cos \left(\tfrac{1}{2}Q_x a\right) \cos \left(\tfrac{1}{2}Q_y a\right)  \\
\lambda_{BC} =& 4S^{-1}[J_1 + 2J_3 \cos Q_ya] \cos \left(\tfrac{1}{2}Q_x a\right) \cos \left(\tfrac{1}{2}Q_z a\right)  \\
\lambda_{AC} =& 4S^{-1}[J_1 + 2J_3 \cos Q_xa] \cos \left(\tfrac{1}{2}Q_y a\right) \cos \left(\tfrac{1}{2}Q_z a\right)  \\
\end{split}
\end{equation}
%\begin{eqnarray}
%\lambda_{AB} &=& 2J \cos((q_x-k_y)\frac{a}{2}) + 2J' \cos((k_x+k_y)\frac{a}{2}) \nonumber \\
%\lambda_{BC} &=& 2J \cos((k_x-k_z)\frac{a}{2}) + 2J' \cos((k_x+k_z)\frac{a}{2}) \nonumber \\
%\lambda_{AC} &=& 2J \cos((k_y-k_z)\frac{a}{2}) + 2J' \cos((k_y+k_z)\frac{a}{2}) 
%\end{eqnarray}
where coordinates are in terms of cube axes with lattice constant $a$. Note that for the spin-wave frequency, the bare exchange and anisotropy constants are divided by $S$.  

In general, numerical analysis is required to obtain the spin wave frequencies as a function of wave vector but some special cases mentioned in I can again be
determined analytically.  For all of the numerical results shown below, values from the DFT calculation given in the Introduction were used. 

\subsection{Zero Anisotropy}
For the case $K=0$,  the eigenvalue problem involves the $ 3 \times 3$ symmetric matrix Eq. (7) in I with elements now given by

%\begin{equation}
%\left( \begin{array}{ccc} A_1 & B_1 & B_2 \\ B_1 & A_2 & B_3 \\ B_2 & B_3 & A_3 \end{array} \right)
%\end{equation}
 
\begin{equation}
\begin{split}
A_1 =& X^2 - (\lambda_{AB}^2 + \lambda_{AC}^2)/2 \\
A_2 =& X^2 - (\lambda_{AB}^2 + \lambda_{BC}^2)/2  \\
A_3 =& X^2 - (\lambda_{AC}^2 + \lambda_{BC}^2)/2  \\
B_1 =& -X\lambda_{AB}/2 - \lambda_{AC}\lambda_{BC}/2 \\
B_2 =& -X\lambda_{AC}/2 - \lambda_{AB}\lambda_{BC}/2 \\
B_3 =& -X\lambda_{BC}/2 - \lambda_{AB}\lambda_{AC}/2
\end{split}
\end{equation}
%\begin{eqnarray}
%A_1 &=& 4(J + J')^2 - (\lambda_{AB}^2 + \lambda_{AC}^2)/2 \nonumber \\
%A_2 &=& 4(J + J')^2 -(\lambda_{AB}^2 + \lambda_{BC}^2)/2 \nonumber \\
%A_3 &=& 4(J+J')^2 -   (\lambda_{AC}^2 + \lambda_{BC}^2)/2 \nonumber \\
%B_1 &=& (J+J')\lambda_{AB} - \lambda_{AC}\lambda_{BC}/2 \nonumber \\
%B_2 &=& (J+J')\lambda_{AC} - \lambda_{AB}\lambda_{BC}/2 \nonumber \\
%B_3 &=& (J+J')\lambda_{BC} - \lambda_{AB}\lambda_{AC}/2 
%\end{eqnarray}

The special case $Q_x=Q_y=Q_z$ again yields eigenvalues of the general form given by Eq. (11) in I but further reduction in terms of the $J_i$ is not illuminating.
However, for the case where the wave vector is directed along a cube axis, $Q_y=Q_z=0$ for example, it can be shown that one of the three modes takes the form
\begin{equation}
\begin{split}
\omega_1^2 = &S^{-2}(2J_2 + 8J_4 - 8J_3)[(6J_1 + 12J_3)(\cos Q_xa -1) \\
           + &(2J_2 + 8J_4 + 4J_3)(\cos Q_xa -1)^2]
\end{split}
\end{equation}
%\begin{eqnarray}
%\omega_1 &=& 0 \nonumber \\
%\omega_2 &=& \omega_3 = 2(J+J')|\sin(k a/2)|
%\end{eqnarray}
This yields the zero energy mode Eq. (13) of I in the absence of further-neighbor exchange.
It is also zero if $2J_2 + 8J_4 = 8J_3$, which is not possible in the case of IrMn$_3$ with J$_3$ $>$ 0 and J$_2$, J$_4$ $<$ 0. 
Thus, as in the 2D case, further-neighbor exchange removes the flat mode.

\subsection{Effects of Anisotropy}

\begin{figure}[htp!]
\centering
\includegraphics[width=1.0\textwidth]{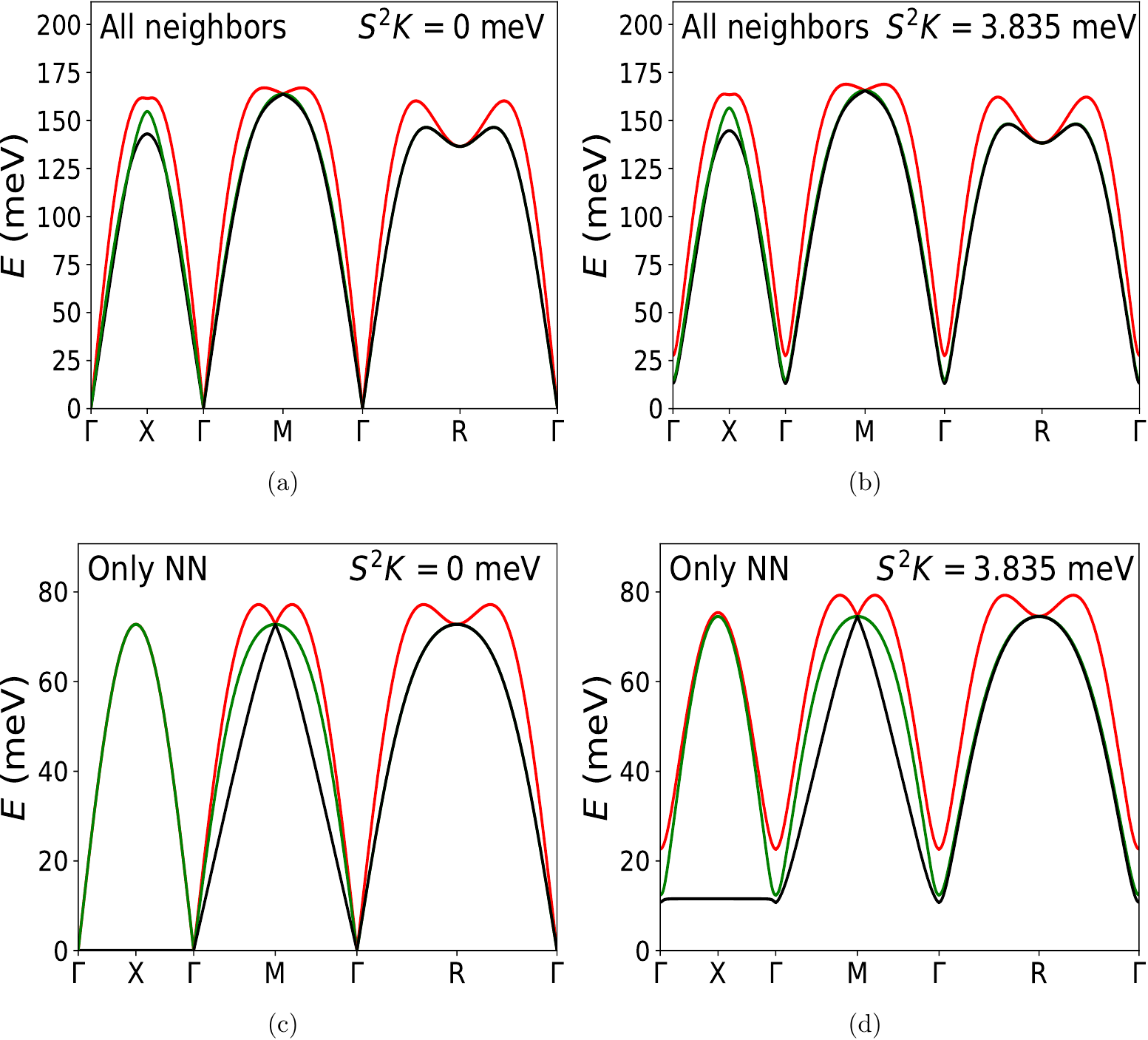}
\caption{Spin wave modes along the $\Gamma$X, $\Gamma$M and $\Gamma$R directions illustrating the impact of anisotropy and further-neighbor exchange interactions.  In the case of NN exchange only with $K$=0 (panel (c)), there is a zero energy mode along $\Gamma$X (black line) and the remaining two modes (red and green lines) are degenerate. Values from the DFT calculation given in the Introduction were used.}\label{wq}
\end{figure}

Fig.~\ref{wq} shows the spin wave frequencies $\omega$ along the $\Gamma$X$(100)$, $\Gamma$M$(110)$ and $\Gamma$R$(111)$ directions with only NN exchange $J_1$ included and also with further-neighbor $J_2$, $J_3$, and $J_4$ added, with and without cubic anisotropy $K$ (also see I).  The effect of further-neighbor interactions is to lift the degereracy that gives rise to the flat zero energy mode (black line) that occurs along $\Gamma$X with only NN interactions (Figs.~\ref{wq}a and c).  The resulting dispersion of this mode is now nearly the same as the other two branches. Note as well that the degeneracy of the other two modes (red and green lines) occuring around the X point in the case of only NN exchange is split with longer-range interactions included.  As before, the impact of  anisotropy $K>0$ is to introduce a gap at the zone center (Figs.~\ref{wq}b and d).  Also note that two of the modes along the $\Gamma$ - M line are degenerate with further-neighbor exchange added (green and black lines in Figs.~\ref{wq}a and b) that were well separated with only NN interactions (Figs.~\ref{wq}c and d).  In contrast, the near degeneracy of these modes around the R point is little impacted by including further-neighbor exchange.

At the zone center $\mathbf{Q}=\mathbf{0}$, the small $K$ dependence of the modes can be calculated and are given by
\begin{equation}
\begin{split}
\omega_1 \simeq \omega_2 \simeq&\ 2S^{-1}\sqrt{(J_1+2J_3)K} \\
\omega_3 \simeq&\ 4S^{-1}\sqrt{(J_1+2J_3)K} 
\end{split}
\end{equation}
%\begin{eqnarray}
%\omega_1&\simeq& \omega_2 \simeq\sqrt{2(J+J')K} \nonumber \\
%\omega_3&\simeq& 2 \sqrt{2(J+J')K}
%\end{eqnarray}
yielding the prediction of small ${\bf Q}$ gaps of about 13 meV and 26 meV, respectively, for IrMn$_3$.

\subsection{Dynamic structure factor}

The Green's function method\cite{MarshallandLovesey1971} used in I was applied here to calculate the part of the dynamic structure factor that contributes to the inelastic neutron scattering cross section 
\begin{equation}
S(\mathbf{Q},\omega)= \sum_{m,n=x,y,z} S^{mn}(\mathbf{Q},\omega)(\delta_{mn} - \hat{Q}_m \hat{Q}_n)
\label{eq:structurefactor}
\end{equation}
where $S^{mn}(\mathbf{Q},\omega)$ is the double Fourier transform of the correlation function $<S_i^m(0) S_j^n(t)>$, to provide  an indication of the inelastic neutron scattering response for IrMn$_3$ with all four NN exchange interactions and anisotropy included.

\begin{figure}[htp!]
\centering
\includegraphics[width=0.9\textwidth]{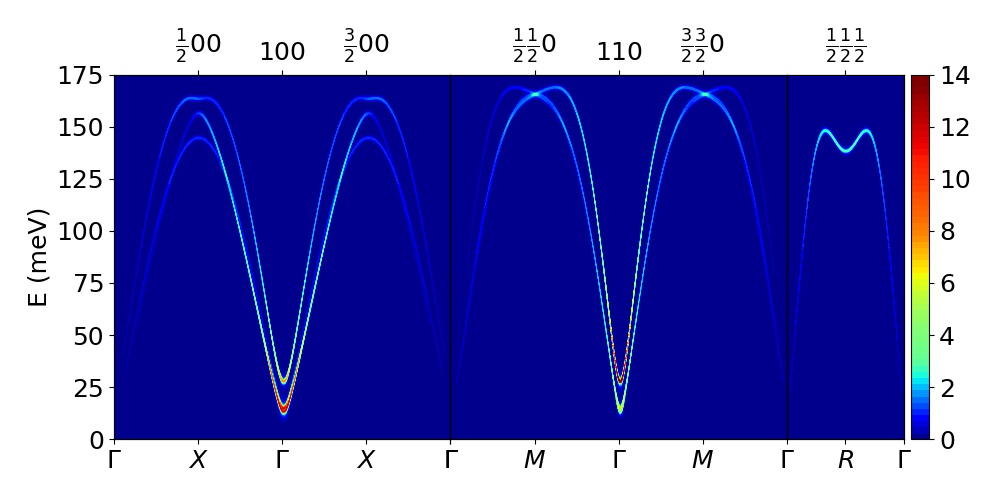}
\caption{Dynamic structure factor, $S(\mathbf{Q},\omega)$, for $\mathrm{IrMn_3}$ assuming a single magnetic (111) domain along the three principal cubic directions. Indicated wave vectors show the symmetric points in the dispersion.
Exchange interactions J$_1$, J$_2$, J$_3$, and J$_4$, as well as anisotropy are included. Values from the DFT calculation given in the Introduction were used. 
%{\color{red} Garrett: Can you add lables $\Gamma$, X, R, M as in Fig. 2?}
}\label{Fig2}
\end{figure}

Fig.~\ref{Fig2} shows $S(\mathbf{Q},\omega)$ assuming a single magnetic $(111)$ domain for $\mathbf{Q}$ along the three principal cube directions with further-neighbor exchange as well as anisotropy included. 
Corresponding results with only NN exchange and $K$ included for $\mathbf{Q}$ along [100] may be found in I.     
Of particular note is that for the cases with $\mathbf{Q}$ along $[100]$ and $[110]$, the intensity is expected to be relatively small in the first Brillouin zone but is substantially larger in the second zone. This is not the case for $\mathbf{Q}$ along $[111]$ but the overall intensity is much weaker for the wave vector in this direction.

Fig.~\ref{Fig2} can be compared with Fig.~\ref{wq}b, illustrating the removal of the low frequency mode along $\Gamma$X with the addition of extra neighbors and the splitting of the degeneracy of the higher frequency modes. Noticeably, along the [111] direction there is a mode not seen in Fig.~\ref{Fig2}c that is present in Fig.~\ref{wq}b. This is not an added degeneracy, but is rather due to a perfect cancellation in intensity when applying Eq.~\ref{eq:structurefactor}. A similar effect is observed in the elastic scattering results of Fig. 3 of I where there is no peak at [111] unless anisotropy is added.

% with $J=J'=1$ for both $K=0$ and $K/J=0.1$. We allow the wave vector to extend beyond the first zone boundary for Fig.~\ref{Fig4}(a) and Fig.~\ref{Fig4}(b), but restrict it to be in the first !zone for Fig.~\ref{Fig4}(c) and Fig.~\ref{Fig4}(d).  In Fig.~\ref{Fig4}(a) and Fig.~\ref{Fig4}(b), the intensity is very large at the wave vectors $\vec{\kappa}=\frac{2\pi}{a}(h,k,l)$ corresponding to the elastic peaks with $h,k,l$ not all even or odd, but the scale is such that the smaller wave vector modes (which can be seen in Fig.~\ref{Fig4}(c) and Fig.~\ref{Fig4}(d)) are not visible due to the large intensity near the elastic peaks located at $\vec{\kappa}=\frac{2\pi}{a}(1,0,0)$. The flat mode is clearly visible in Fig.~\ref{Fig4}(d) on the smaller scale but has an intensity much reduced from the other modes at the zone boundary. Fig.~\ref{Fig4} can be compared with Figs.~\ref{Fig2}(e) and (f), illustrating the appearance of the low frequency mode along $\Gamma$X and the splitting of the degeneracy of the higher frequency modes. While the $[111]$ direction ($\Gamma$R) is not shown, the intensity is on the order of a hundred times smaller than the [100] direction.

\section{INELASTIC NEUTRON SCATTERING}

\subsection{Model for scattering cross section}

To provide a more meaningful comparison with the experimental data, the relevant parts of the scattering cross section for inelastic magnetic scattering are also calculated \cite{MarshallandLovesey1971}
\begin{equation}
 \left(\frac{d^2\sigma}{d\Omega dE'}\right)_{\mathrm{inel}} \propto \frac{k'}{k} |F(\mathbf{Q})|^2 S(\mathbf{Q},\omega),
\label{eq:inelasticprop}
\end{equation}
where $S(\mathbf{Q},\omega)$ is given by Eq.~\ref{eq:structurefactor} and $F(\mathbf{Q})$ is the magnetic form factor.
Additional kinematic factors in the cross section, e.g., $\frac{k'}{k}$ (where $k$ and $k'$ are the initial and final scattering wavevector magnitudes, respectively) are accounted for in the reduction of the experimental data.
This form factor is given by, in the usual dipole, or spherical, approximation,\cite{haraldsen2012,brownwebsite}
\begin{equation}
\label{eq:magneticformfactor}
 F(s) = Ae^{-a s^2}+Be^{-b s^2}+Ce^{-c s^2} + D,
\end{equation}
where $s = Q/4\pi$. This function approaches zero at $s \gtrsim 1$\AA$^{-1}$ and is negligible for $Q \gtrsim 4\pi$\AA$^{-1}$.
In principle the values of the dimensionless constants depend on the oxidation state of Mn. However, little variation is observed so the tabulated values for Mn0 were used.\cite{brownwebsite} 
The wavevector is calculated through $Q = \frac{2\pi}{a}\sqrt{h^2+k^2+l^2}$, using the low temperature lattice constant $a =3.76(1)$. \cite{tomeno1999} 
%As the oxidation state of Mn does not have a large impact on the form factor, Mn0 was chosen in the calculations that follow.

\subsection{Calculations for comparison with neutron scattering data}

For comparison of the calculated cross section with the data from the polycrystalline sample used in the SEQUOIA experiment, $S(\bf{Q},\omega)$ is averaged over all crystallographic directions, yielding 
\begin{equation}
 S(Q,\omega) = \int_{\phi = 0}^{2\pi}\int_{\theta = 0}^{\pi} S(\bf{Q},\omega)\sin\theta \, d \theta \, d \phi,
\end{equation}
where $\theta$ and $\phi$ are the azimuthal and polar angles describing the orientation of ${\bf Q}$.  This integral can be approximated through Monte Carlo integration using the following expression: 
\begin{equation}
 S(Q,\omega) = \frac{1}{\mathrm{n}}\sum_{i=1}^{\mathrm{n}} S(Q,\arccos(2a_i-1),2\pi b_i,\omega).
\end{equation}
In the equation above, $a$ and $b$ represent two random numbers from 0 to 1 and $n$ is the number of iterations. Typically, 1000 random directions are chosen for each Q. For the model results presented below, the scattering cross section Eq.~\ref{eq:inelasticprop} was then calculated.

\begin{figure}[htp!]
\centering
\includegraphics[width=0.95\textwidth]{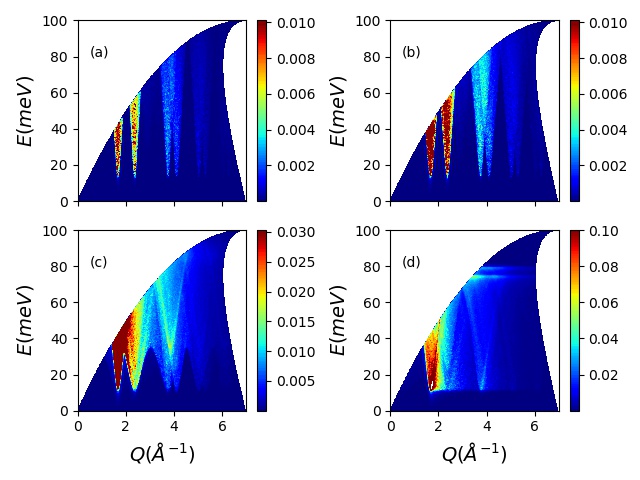}
\caption{Calculated, powder-averaged inelastic neutron scattering spectrum for E$_i$~$=$~100 meV with (a) all four further-neighbor exchange J$_1$, J$_2$, J$_3$, and J$_4$; (b) with J$_1$, J$_2$, and J$_3$ only; (c) with J$_1$, and J$_2$ only; and (d) with nearest neighbor exchange J$_1$ only.
Anisotropy is included in all cases and these model results suggest a gap of about 13 meV. Values from the DFT calculation given in the Introduction were used.}\label{Fig5}

%(a), (b) Calculated, powder-averaged inelastic neutron scattering spectrum for $E_i$~$=$~100 meV with further-neighbor exchange J$_1$, J$_2$, J$_3$, and J$_4$, as well as anisotropy included.\new{Values from the DFT calculation given in the Introduction were used.} The same simulation is shown in panels (a) and (b), but the intensity scale is different in the two cases. Panel (c) shows calculated results using only NN exchange J$_1$  with anisotropy. Panel (d) shows the case with all of the further-neighbor interactions included but without anisotropy. In the case of further-neighbor exchange, the higher intensity plot presented in panel (a) shows the strongest, nearly-vertical columns of scattering centered above (100), (110), and (210) magnetic Bragg peak positions for IrMn$_3$, corresponding to $Q$~$=$~1.67 \AA$^{-1}$, 2.37 \AA$^{-1}$ and 3.75 \AA$^{-1}$, respectively. For the case of NN exchange only, the calculation shown in panel (c) reveals that the magnetic scattering centered above $Q$~$=$~2.37~\AA$^{-1}$ and 3.75~\AA$^{-1}$ should be comparable in intensity. However, this feature is not observed in the SEQUOIA data.  }\label{Fig5}
\end{figure}

Figure \ref{Fig5} shows the calculated powder-averaged spectra, for comparison to the inelastic scattering results acquired with $E_i$~$=$~100 meV, for four cases of setting selected further neighbor exchange constants equal to zero.  All of the results include anisotropy. The energy gap appears to be about 13 meV from these model results.

The strongest intensity features predicted by the model that includes further-neighbor exchange up to at least $J_3$ (Figs.~\ref{Fig5}a and b) at Q=1.67, 2.37,  3.75 and 4.10 \AA$^{-1}$~ are observed.  This is not surprising as they are above the (100), (110), (210), and (211) magnetic Bragg peak positions.
Figure \ref{Fig5}c shows that these excitations are less distinct from each other indicating that $J_3$ is important to provide the overall bandwidth of the excitation spectrum.
 A significant difference between the NN only model (Fig.~\ref{Fig5}d) and the others with further-neighbor exchange is that the  inelastic feature above Q=2.37 \AA$^{-1}$~(110 peak position) is much more intense and well-defined with the inclusion of $J_3$ and $J_4$.  
 The other regions of well defined but lower intensity resulting from the further-neighbor model at larger wave vectors also do not occur within the NN only model.

%ADAM: WE SHOULD THINK ABOUT INCLUDING FIG 6 (see powderAll2.pdf) 
\begin{figure}
\includegraphics[width=0.75\textwidth]{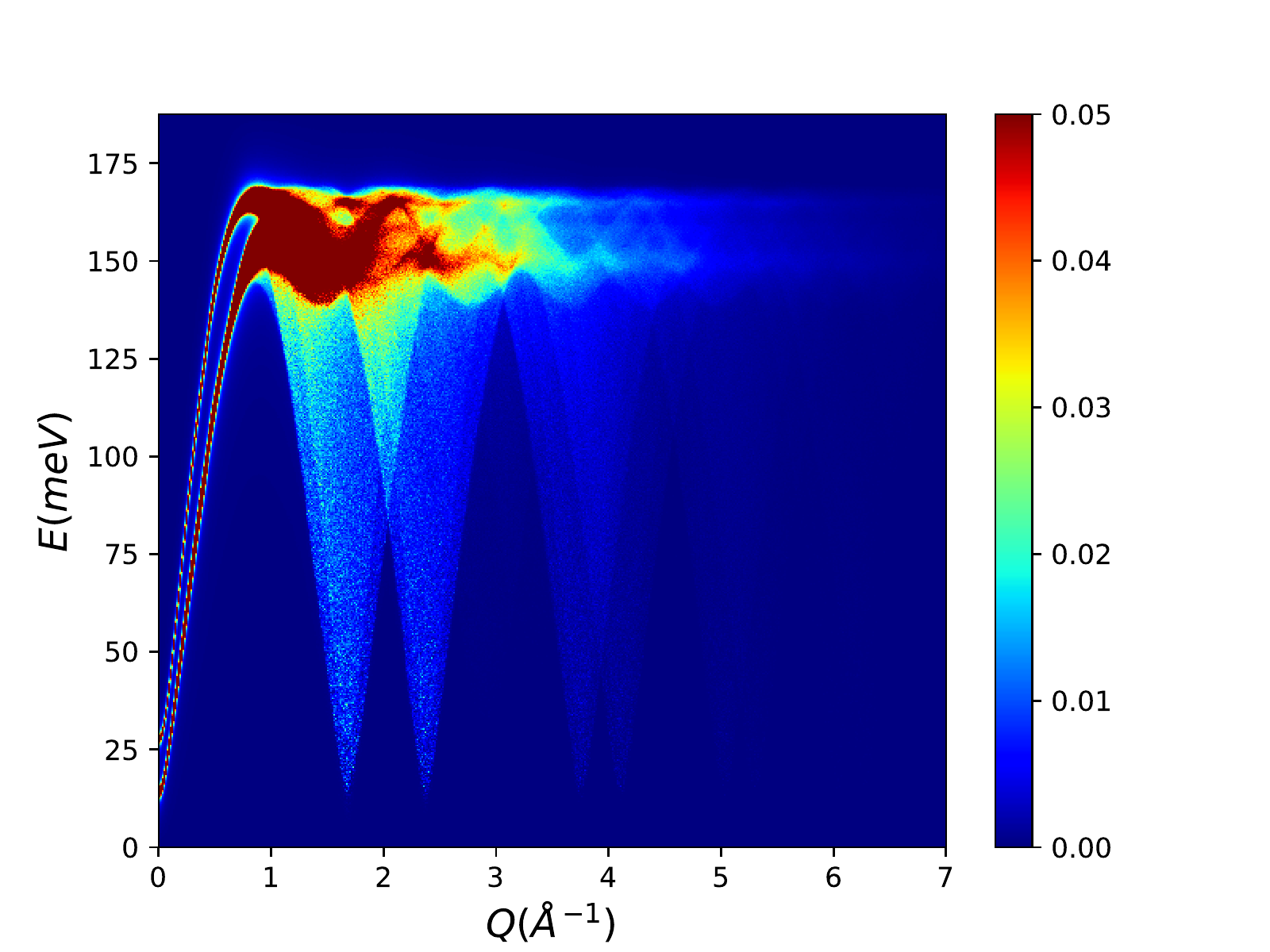}
\caption{The powder averaged calculation showing the full range of the dispersion for the model including anisotropy and all of the neighbor interactions.  Values from the DFT calculation given in the Introduction were used.}\label{Fig6}
\end{figure}

As shown below, a comparison of the experimental data with theses figures show that the inclusion of all four exchange interactions yield the best agreement.
In Figure \ref{Fig5} the kinematic constraints of the neutrons scattering experiment were applied to allow for comparison. 
In Figure \ref{Fig6} that constraint is removed to be able to compare to multiple $E_i$ values.  In other words, it shows the full powder averaged dispersion with the contributions from all of the neighbors.
The data in this figure can then be used as input in the Monte Carlo ray tracing simulations, that account for instrumental effects such as resolution and are described below, for closer comparison to the measurements.

\subsection{Sample preparation}

Mn and Ir powders in the molar ratio of 3:1 were mixed well and pelletized. The pellets were then sealed inside of a quartz tube under approximately 1/3 atmosphere of argon gas. The sealed ampoule was heated to 1050$^\circ$C in 10 hours and kept at this temperature for 48 hours before cooling to 600$^\circ$C in 6 hours. After dwelling at 600$^\circ$C for 7 days, the ampoule was quenched into iced water. X-ray diffracton confirmed that the resulting pellet was nearly single phase IrMn$_3$ (ordered Mn and Ir,  {\it Pm3m} space group), with a small amount of MnO detected on the surface. No preferred crystal orientation was detected.

\subsection{Inelastic neutron scattering: experimental details}

Inelastic neutron scattering experiments were performed on the direct-geometry time-of-flight (TOF) chopper spectrometer SEQUOIA\cite{Granroth2006,Granroth2010} at Oak Ridge National Laboratory's Spallation Neutron Source. 
The polycrystalline sample was cut into thin slabs to minimize neutron absorption in the INS experiment. 
The cell was then cooled down to 5~K in a closed cycle refrigerator.
Spectra were collected with incident energies $E_i$~$=$~50, 100, 300, and 500 meV in coarse energy resolution mode (elastic resolution of $\sim$ 4 \% $E_i$) to investigate the spin wave excitations in the magnetically-ordered phase. 
An empty Al cell was measured in identical experimental conditions for the $E_i$ = 50, 100, and 500 meV cases  and the resulting spectra were subtracted from the corresponding sample spectra.
Additional $E_i$ = 300 meV data was also collected, but since this data was only used to investigate magnetic excitations above the cutoff for the Al phonon density-of-states the empty Al cell was not measured in this case.
Coarse energy resolution mode was achieved using Fermi Chopper 2 operating at 240~Hz ($E_i$~$=$~50 or 100 meV), 480 Hz ($E_i= 300$ meV) or 600 Hz ($E_i$~$=$~500~meV), and the background from the prompt pulse was removed with a $T\rm{_{o}}$ chopper operating at 60~Hz ($E_i$~$=$~50 or 100 meV), 120 Hz ($E_i= 300$ meV) or 150 Hz ($E_i$~$=$~500~meV).\cite{Granroth2010}

\begin{figure}[htp!]
\centering
\includegraphics[width=0.95\textwidth]{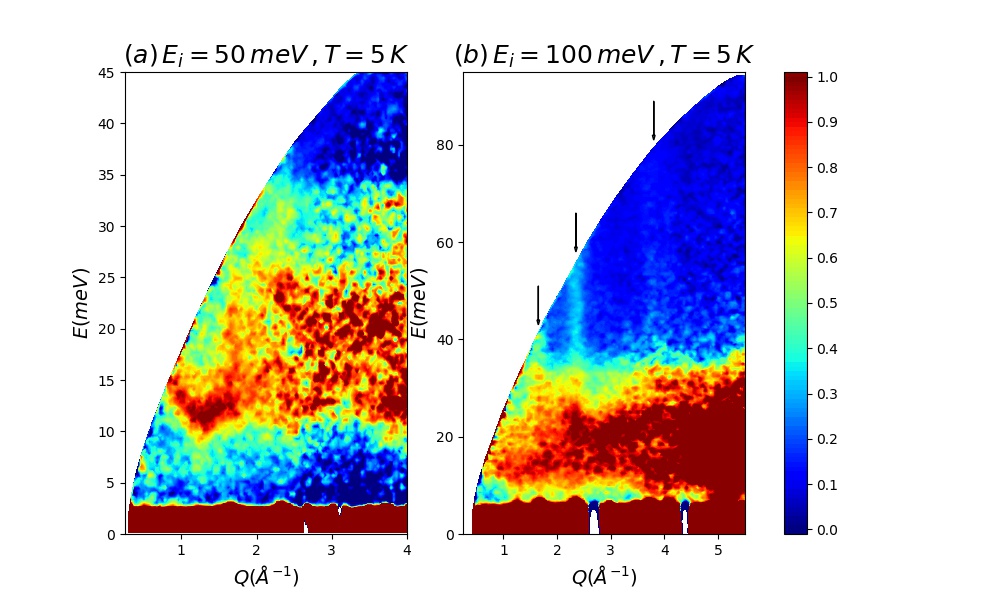}

\caption{(a) Color contour plots of spectra measured on SEQUOIA. 
(a) For $E_i$~$=$50~meV we observe a magnon band due to a small MnO impurity, a nearly-vertical column of magnetic scattering centered about $Q$~$=$~1.67~\AA$^{-1}$  corresponding to the magnon band for $\mathrm{IrMn_3}$, and some phonon modes. 
%(b) A constant-$Q$ cut through the data presented in (a), with an integration range of $Q$~$=$~[1.5,1.9]~\AA$^{-1}$. 
(b)  For $E_i$~$=$~100~meV we observe
 four nearly-vertical columns of scattering correspond to IrMn$_3$ spin wave excitations. 
%(d) A constant-$E$ cut through the data presented in (c), with an integration range of $E$~$=$~[35, 45] meV. 
%(c) Data at $E_i$~$=$~500~meV shows some higher energy scattering not accounted for by the local spin-wave model. 
%(f) Two constant-$Q$ cuts through the data presented in (e), with integration ranges of $Q$~$=$~[3.5, 4.5]~\AA$^{-1}$ and [6, 7]~\AA$^{-1}$. The broad peak observed in the lower $Q$-cut corresponds to the top of the magnon band for IrMn$_3$.
 }
\label{Fig4}
\end{figure}

\subsection{Inelastic neutron scattering: results}
The inelastic neutron results are reduced from TOF and pixel position to $Q$ and $E$ using Mantid \cite{ARNOLD2014156} and the $S(Q,E)$ slices and cuts were generated from this reduced data using DAVE.\cite{azuah2009dave}
The color contour plots in Fig.~\ref{Fig4} summarize the main results from the SEQUOIA experiment. Panels (a), and (b) depict the IrMn$_3$ spectra collected with $E_i$= 50 and 100 meV, respectively.  
As $E_i$ is increased, the energy resolution coarsens,  and the kinematic range broadens. 
Several prominent features are shown in the figure. First and most interesting are the three nearly-vertical columns of scattering indicated by the arrows in Fig. \ref{Fig4}b.  
These columns are centered above $Q$~$=$~1.67, 2.36, 3.75, and 4.10~\AA$^{-1}$, which correspond to the (100), (110), and  the combined (210) and (211) magnetic Bragg peak positions for IrMn$_3$.
A more detailed examination of these excitations show that they match the further-neighbor model calculations as will be discussed in detail below. Second, there is a strong phonon background  between $\sim 10 - 35$ meV .  
Third,  the weakly-dispersive magnetic mode, between 10 and 18 meV in a $Q$ range of 0.7-2.5~\AA$^{-1}$, can be attributed to a magnon band arising from  $2\pm1\%$ of MnO impurities.  

%This value is best quantified by examining the constant-$Q$ cuts shown in Fig.~\ref{Fig4}(f). Two different cuts, with integration ranges of $Q$~$=$~[3.5, 4.5]~\AA$^{-1}$~and [6, 7]~\AA~$^{-1}$ are compared here. A broad, weak peak centered around 175 - 200 meV is clearly observed in the lower-$Q$ cut, which defines the energy scale for the top of the band. The extremely steep dispersion is consistent with the high ordering temperature for this material.

\subsection{Comparison between SEQUOIA data and spin wave calculations}

To perform a quantitative comparison to the theoretical calculations,  the results were fed into a McVine Monte Carlo ray tracing simulation of the SEQUOIA instrument.\cite{Lin2016}
The simulation  provides a computer model of each component of the instrument including the sample.
Simulated neutrons are then propagated through the model. 
Once a sufficient number are propagated, the calculated result is analyzed in the same way as the data. 
The end result is a directly comparable calculation with the instrumental resolution implicitly included.  
McVine provides a straightforward way to model multiple contributions like impurity phases and multiple scattering.
Both the excitations from the IrMn$_3$ and the MnO were modeled.  However, the unknown phonon contribution  was not included in the simulation.  
A few simulations with multiple scattering were tried but did not improve the agreement
with the measured data so the given results show single scattering events only. 
For the simulations, $1\times10^{10}$ neutron probability packets were propagated from the source and  through the various instrument components.

\begin{figure}
\includegraphics[width=0.5\textwidth]{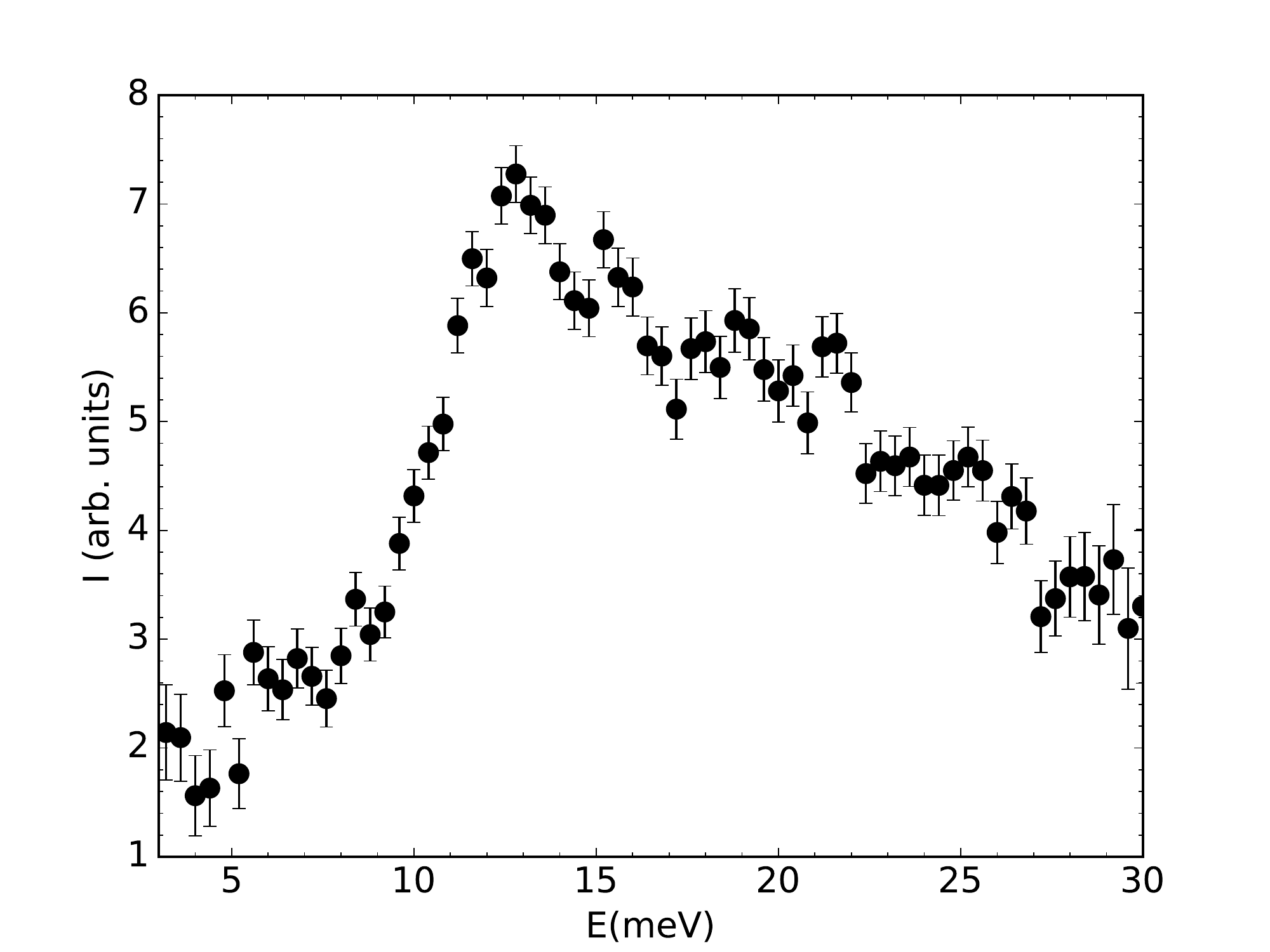}
\caption{A constant-$Q$ cut through the data presented in Figure \ref{Fig4}a, with an integration range of $Q$~$=$~[1.5,1.9]~\AA$^{-1}$. }
\label{cut_100}
\end{figure}

Figure \ref{cut_100} shows a  constant-$Q$ cut of the experimental data, with an integration range of Q = 1.5- 1.9 \AA$^{-1}$ This $Q$ range is centered around the (100) Bragg peak which has both structural and magnetic components due to the known $q = 0$ magnetic order. \cite{tomeno1999}  The mode appears to be strongly-gapped, as the spectral weight decreases significantly at energy transfers below the MnO magnon band.  These measurements estimate the value of the gap to be $12\pm2$ meV, a value  in good agreement with the model results (based on DFT calculations of the cubic anisotropy)  and the McVine simulations. Further refinement of the gap value is hampered by the phonon background.

\begin{figure}
\centering
\includegraphics[width=0.75\textwidth]{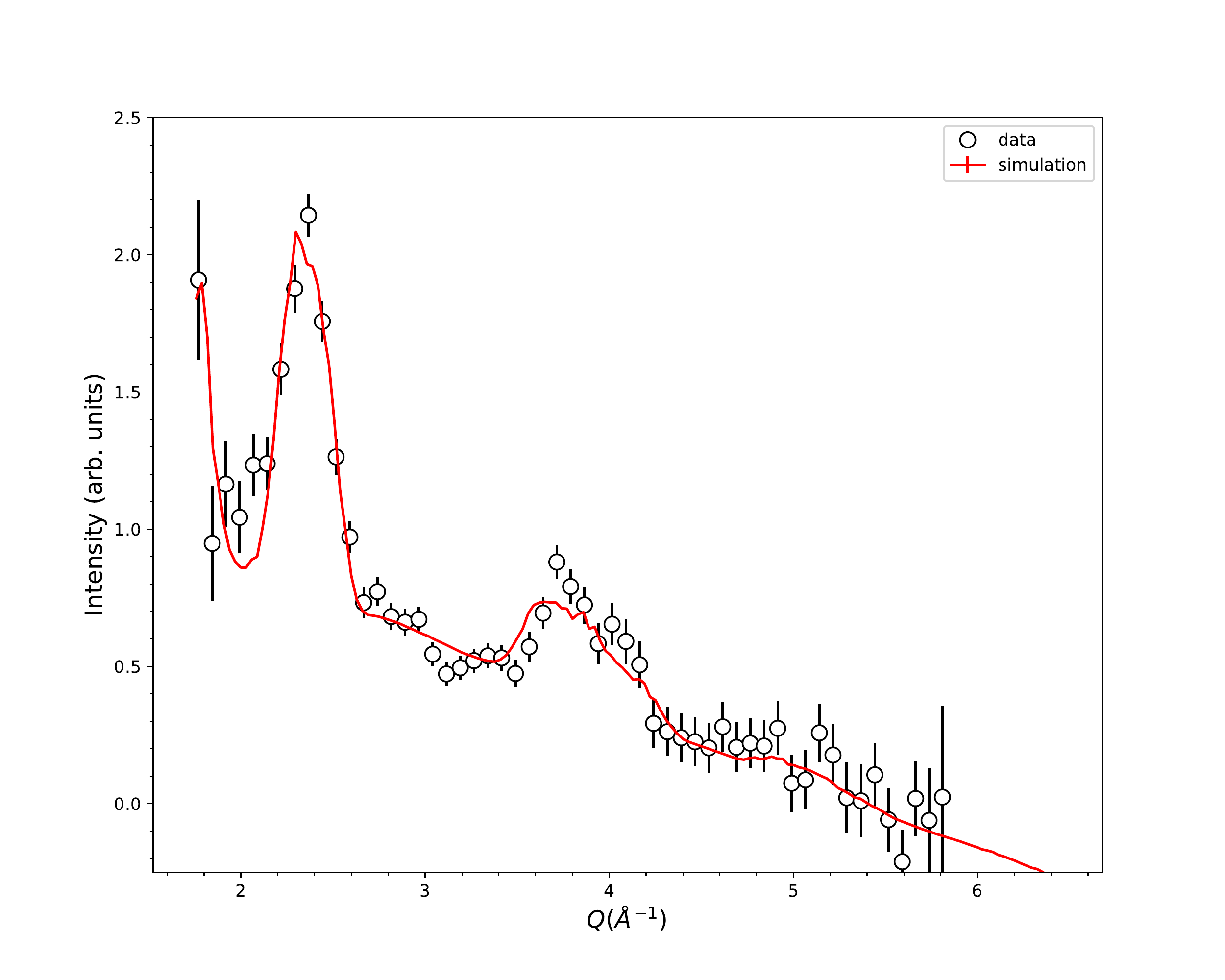}
\caption{A constant $E$ cut of the $E_i$=100 meV data shown in \ref{Fig4}b is indicated by the black circles.  The integration range is between 40 and 70 meV.  
The same analysis procedure was carried out on results from McVine Monte Carlo ray tracing simulations that included the theoretical model for the sample. 
The results from the simulation are given by the red line. The error bars for the simulations arise from statistical uncertainty in the Monte Carlo method and are significantly less than
those of the measurement.}
\label{cut_100meV}
\end{figure}

As  mentioned earlier, for the $E_i$ = 100 meV measurements, the $Q$=2.36 \AA$^{-1}$ peak is clearly observed in Fig.~\ref{Fig4}b.  
However for the calculations, it is not clearly observed in the NN model (Fig.~\ref{Fig5}c) and only becomes prominent when further-neighbor interactions are included (Fig.~\ref{Fig5}b).
It can thus be concluded that a model with NN exchange only does not explain the measured data.
Further examination of  Fig. \ref{Fig4}b shows that the phonon background is greatly reduced above 40 meV, which ensures that there is a large region of Q and E space with magnetic contributions only.
% This allows for a straightforward and direct comparison with the models.  
This provides a region in $Q$ and $E$ space that is cleanly magnetic and facilitates a straightforward and direct comparison with the models. 
To this end, a constant $E$-cut with an integration range of 40 - 70 meV was considered.  
The result is shown as the black circles in Fig. \ref{cut_100meV}. 
The same cut was taken from the $E_i$ = 100 meV McVine simulations and is shown as the red line superimposed on that data. Note that the simulation includes a small linear background contribution. 
 The simulation accounts for the data quite nicely and quantitatively shows that the theoretical model is appropriate.
If the calculated dispersion curves were significantly different it would show up in the widths of the peaks.  
 %%% further discussion of impurity contribution %%%
%We used the elastic channel ($E$~$=$~[-4, 4] meV) of the $E_i$~$=$~100~meV dataset to estimate the volume fraction of the impurity by comparing the observed integrated intensities of the (321) Bragg peak for IrMn$_3$ and the (311) Bragg peak for MnO, and we found a value of 1\%. As a separate check the percentage of MmO was increased in the simulation until the ratio of the IrMn_3 excitations and the MnO excitations matched the ratio in the measurement. These two methods give us a resultof $2\pm1\%$ of MnO impurities

\begin{figure}
    \centering
    \includegraphics[width=0.75\textwidth]{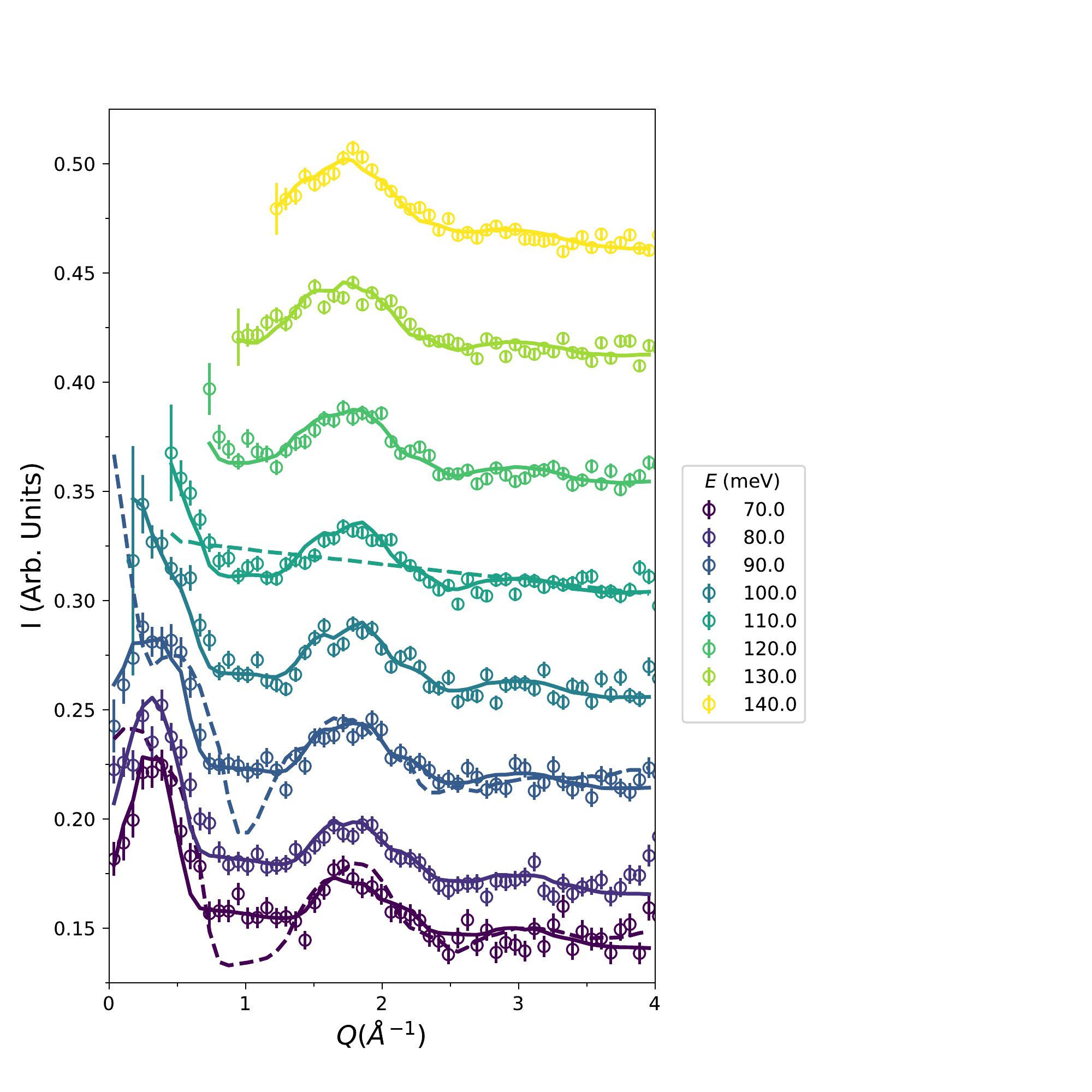}
    \caption{The points are series of cuts from the $E_i$ = 300 meV data. 
    The solid curves are simulations using the described dispersion with all four exchange interactions included.
    %{\color{red} Garrett - Add a broken line or two corresponding to your results with $J_4$ = 0?}
    Broken lines are from a model with $J_4 = 0$. 
    There is a curve, and its corresponding simulation, for each incident energy which are offset for clarity.  
    The same scale factor is applied to all of the curves and an individual sloping background is fit to account for varying backgrounds. 
   Only three curves of the $J_4$ = 0 case are shown as they are sufficient to illustrate that inclusion of J$_4$ results in superior agreement with the data.  }
    \label{cuts_300meV}
\end{figure}

For a broader comparison, data was acquired at $E_i$ 300 meV. 
This setting allowed the data to be acquired with sufficient range and resolution to compare to the model over the 70 - 140 meV range.  
For this comparison 10 meV wide cuts were taken in 10 meV steps.  
The data are shown as points in Figure \ref{cuts_300meV} and the simulations are superimposed on the data with lines of the same color.
As mentioned above, a linear background was added to the simulation and an overall scale factor was also incorporated to allow for a direct comparison to the SEQUOIA data. 
There were no other free parameters in the fit. 
Note the excellent agreement between the model and the data confirming that the additional neighbor model accounts for the magnetic excitations observed in the measurements. 
A similar process of fitting the simulation with a scale factor and a linear background was tried for a case where $J_4 = 0$. 
The results are shown as the dashed curves in Figure \ref{cuts_300meV}. 
Note that at low energies this model does not reproduce the data and furthermore results in a narrower bandwidth of excitations yielding a flat line at 110 meV.  

\begin{figure}
\centering
\includegraphics[width=1.0\textwidth]{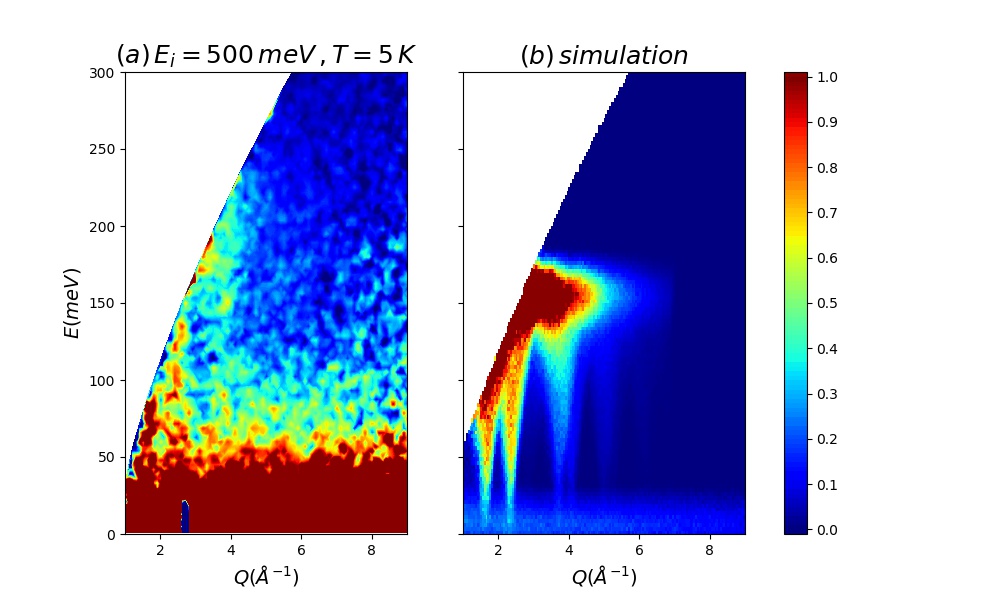} 
\caption{Data (a) and simulation (b) of the $E_i=500$ meV configuration. 
The results indicate that the measurement shows additional scattering above the top of the band as compared to the model.  Also, the high-energy magnetic scattering seems to decrease with $Q$ faster in the experimental data as compared to the simulation. }
\label{sqw_500meV}

\end{figure}

Finally, measurements with an $E_i$ of 500meV were used to capture the top of the magnon band.
Figure \ref{sqw_500meV}  shows the results of the spin wave model and the corresponding simulation. 
The measurements reveal magnetic scattering higher in $E$ than is described by the simulation. 
There are two possibilities to explain this excess scattering. 
First, the exchange parameters as predicted by the DFT calculations could be too low.  Second, there could be another excitation above the magnetic excitations described above.  
Close examination of spin-dependent DFT calculations in Chen {\it et al.} \cite{PhysRevLett.112.017205} indicates there might be  some nested orbitals around the M point which could be consistent with the observed scattering.  
Nevertheless, further DFT calculations and measurements with single crystal samples will be needed to definitively distinguish these two possibilities.
%However it is not clear if those bands carry spin, which is necessary for observation by neutrons.   \hl{Still working on additional checks that would bolster the claim that this is not the top of the magnetic band but has some other origin.}  

\section{SUMMARY AND CONCLUSIONS}

The geometry of intersecting $\{111\}$ kagome planes describing the fcc kagome lattice offers a rare example to study a truly 3D frustrated kagome antiferromagnet, as exemplified by IrMn$_3$.
The so-called q=0 triangular spin structure is preserved in the 3D system along with some of the degeneracies of the NN Hesienberg model which stimulated previous work in the 2D case.
A focus of the present results is to quantify the role of further-neighbor exchange interactions within a local-moment model of spin wave excitations in IrMn$_3$, extending our earlier work that involved only NN exchange.\cite{leblanc2014}.
Earlier DFT calculations\cite{szunyogh2009,szunyogh2011} indicate that exchange up to fourth nearest neighbors are substantial in this transition-metal ion conductor that has been widely used as the exchange pinning
antiferromagnetic thin film which is integral to spin valve technology.  The model Hamiltonian also includes a cubic anisotropy term, which is predicted to be strong from
the DFT theory, and induces a gap in the model spectrum.

New inelastic neutron scattering data on powder samples is also presented and compared with corresponding results from the model. The comparison is facilitated through the use of ray tracing simulations that integrate model predictions with instrumental resolution as well as additional features of the sample, which in this case includes a small amount of magnetic scattering from a MnO impurity.\cite{Lin2016} There is also a substantial amount of IrMn$_3$ phonon scattering in our data that was not included in the simulations. 

Model calculations show that the presence of a strong inelastic feature above the (110) magnetic peak, which is observed in the SEQUOIA data, is only consistent with a model including further neighbor interactions. 
This fact provides strong support for the conclusion that further-neighbor interactions are required to successfully model the spin wave spectrum of IrMn$_3$.
Good agreement between the model and data in the case of constant energy cuts supports the conclusion that the DFT estimates of the exchange interactions are reliable. 

The present work provides the first experimental evidence for the existence of the DFT prediction of strong cubic anisotropy in IrMn$_3$ through the good agreement between the spin-wave model
 and a gap observed in the inelastic spectrum.  This is an important result as our earlier work demonstrated that anisotropy induces a uniform magnetic moment along [111]
 directions.\cite{leblanc2013}  This likely indicates an important mechanism for exchange coupling to adjacent ferromagnetic thin films.  Such a moment could also be employed to induce a single magnetic domain in field cooled single-crystal samples.
 
 Evidence also exists in the data for broad high energy (at and above 200 meV) magnetic scattering that does not appear in the local-moment model.
 This feature deserves further investigation in view of the possibility it is associated with the calculated band structure supporting the anomolous hall effect in such non-collinear antiferromagnets. \cite{PhysRevLett.112.017205}

\section{ACKNOWLEDGMENTS}
This work was supported by the Natural Sciences and Engineering Council (NSERC) of Canada, the Canada Foundation for Innovation (CFI), and Compute Canada. A portion of this research used resources at the Spallation Neutron Source, a DOE Office of Science User Facility operated by Oak Ridge National Laboratory(ORNL). The Monte Carlo Ray Tracing simulations were performed on the CADES cloud computing resource at ORNL.


\begin{thebibliography}{}
\bibitem{hemmati2012} V. Hemmati, M. L. Plumer, J. P. Whitehead, and B. W. Southern, Phys. Rev. B \textbf{86}, 104419 (2012).
\bibitem{chalker1992} J. T. Chalker, P. C. W. Holdsworth, and E. F. Shender, Phys. Rev. Lett. \textbf{68}, 855 (1992).
\bibitem{harris1992} A. B. Harris, C. Kallin, and A. J. Berlinsky, Phys. Rev. B \textbf{45}, 2899 (1992).
\bibitem{schnabel2012} S. Schnabel and D. P. Landau, Phys. Rev. B \textbf{86}, 014413 (2012).
\bibitem{matan06} K. Matan, D. Grohol, D. G. Nocera, T. Yildirim, A. B. Harris, S. H. Lee, S. E. Nagler, and Y. S. Lee, Phys. Rev. Lett. \textbf{96}, 247201 (2006).
%\bibitem{syan2011} S. Yan, D. A. Huse, S. R. White, Science \textbf{332}, 1173 (2011).
%\bibitem{rastelli86} E. Rastelli and A. Tassi, J. Phys. C: Solid State Phys. \textbf{19}, L423 (1986); {\it ibid} \textbf{21}, 1003 (1988).
%\bibitem{jansen86} A. P. J. Jansen, Phys. Rev. B \textbf{33}, 6352 (1986).
%\bibitem{bramwell2001} S. T. Bramwell and M. J. P. Gingras, Science \textbf{294}, 1495 (2001).
%\bibitem{zhitomirsky2012} M. E. Zhitomirsky, M. V. Gvozdikova, P. C. W. Holdsworth, and R. Moessner, Phys. Rev. Lett. \textbf{109}, 077204 (2012).
%\bibitem{wong2013} A. W. C. Wong, Z. Hao, and M. J. P. Gingras, Phys. Rev. B \textbf{88}, 144402 (2013).
%\bibitem{ross14} K. A. Ross, Y. Qiu, J. R. D. Copley, H. A. Dabkowska, and B. D. Gaulin, Phys. Rev. Lett. \textbf{112}, 057201 (2014).%arXiv:1401.1176v1(2014).
%\bibitem{villain80} J. Villain, R. Bidaux, J. -P. Carton, and R. Conte, J. Phys. France \textbf{41}, 1263 (1980).
%\bibitem{henley89} C. L. Henley, Phys. Rev. Lett. \textbf{62}, 2056 (1989).
%\bibitem{shahbazi08} F. Shahbazi and S. Mortezapour, Phys. Rev. B \textbf{77}, 214420 (2008).

%\bibitem{leblanc2013} M. D. LeBlanc, M. L. Plumer, J. P. Whitehead, and B. W. Southern, Phys. Rev. B \textbf{88}, 094406 (2013). 
%\bibitem{leblanc2014} M. D. LeBlanc, B. W. Southern, M. L. Plumer, and J. P. Whitehead, Phys. Rev. B \textbf{90}, 144403 (2014).
  
%\bibitem{szunyogh2009} L. Szunyogh, B. Lazarovits, L. Udvardi, J. Jackson, and U. Nowak, Phys. Rev. B \textbf{79}, 020403(R) (2009).
%\bibitem{berk1999} A. E. Berkowitz and K. Takano, J. Magn. Magn. Mater. \textbf{200}, 552 (1999); R. L. Stamps, J. Phys. D \textbf{33}, R247 (2000); M. Blamire and B. Hickey, Nat. Mater. \textbf{5}, 87 (2006).
\bibitem{ogrady2010} K. O'Grady, L. E. Fernandez-Outon, and G. Vallejo-Fernandez, J. Magn. Magn. Mater. \textbf{322}, 883 (2010).
\bibitem{tsunoda10} M. Tsunoda, H. Takahashi, T. Nakamura, C. Mitsumata, S. Isogami, and M. Takahashi, Appl. Phys. Lett. \textbf{97}, 072501 (2010).
%\bibitem{kren66} Kr\'{e}n, G. K\'{a}d\'{a}r, L. P\'{a}l, J. S\'{o}lyom, and P. Szab\'{o}, Phys. Lett. \textbf{20}, 331 (1966);
%E. Kr\'{e}n, G. K\'{a}d\'{a}r, L. P\'{a}l, J. S\'{o}lyom, P. Szab\'{o}, and T. Tarn\'{o}czi, Phys. Rev. \textbf{171}, 574 (1968);
%A. Sakuma, R. Y. Umetsu, and K. Fukamichi, Phys. Rev. B \textbf{66}, 014432 (2002); T. Ikeda and Y. Tsunoda, J. Phys. Soc. Jpn. \textbf{72}, 2614 (2003).
%\bibitem{tomeno1999} I. Tomeno, H. N. Fuke, H. Iwasaki, M. Sahashi, and Y. Tsunoda, J. Appl. Phys. \textbf{86}, 3853 (1999).

%\bibitem{chen2014} H. Chen, Q. Niu, and A. H. MacDonald, Phys. Rev. Lett. \textbf{112}, 017205 (2014).
\bibitem{kohn2013} A. Kohn, A. Kovacs, R. Fan, G. J. McIntyre, R. C. C. Ward, and J. P. Goff, Sci. Rep. \textbf{3}, 2412 (2013).
\bibitem{yanes2013} R. Yanes, J. Jackson, L. Udvardi, L. Szunyogh, and U. Nowak, Phys. Rev. Lett. \textbf{111}, 217202 (2013).
\bibitem{szunyogh2009} L. Szunyogh, B. Lazarovits, L. Udvardi, J. Jackson, and U. Nowak, Phys. Rev. B \textbf{79}, 020403(R) (2009).
\bibitem{szunyogh2017} L. Szunyogh and U. Nowak, private communication (2017).
\bibitem{szunyogh2011} L. Szunyogh, L. Udvardi, J. Jackson, U. Nowak, and R. Chantrell, Phys. Rev. B. \textbf{83}, 024401 (2011).
\bibitem{leblanc2013} M. D. LeBlanc, M. L. Plumer, J. P. Whitehead, and B. W. Southern, Phys. Rev. B \textbf{88}, 094406 (2013). 
\bibitem{leblanc2014} M. D. LeBlanc, B. W. Southern, M. L. Plumer, and J. P. Whitehead, Phys. Rev. B \textbf{90}, 144403 (2014).
\bibitem{yerzhakov2016}  H. V. Yerzhakov, M. L. Plumer, and J. P. Whitehead, J. Phys.: Condens. Matt. \textbf{28}, 196003 (2016).
\bibitem{kren66} Kr\'{e}n, G. K\'{a}d\'{a}r, L. P\'{a}l, J. S\'{o}lyom, and P. Szab\'{o}, Phys. Lett. \textbf{20}, 331 (1966);
E. Kr\'{e}n, G. K\'{a}d\'{a}r, L. P\'{a}l, J. S\'{o}lyom, P. Szab\'{o}, and T. Tarn\'{o}czi, Phys. Rev. \textbf{171}, 574 (1968);
A. Sakuma, R. Y. Umetsu, and K. Fukamichi, Phys. Rev. B \textbf{66}, 014432 (2002); T. Ikeda and Y. Tsunoda, J. Phys. Soc. Jpn. \textbf{72}, 2614 (2003).

\bibitem{tomeno1999} I. Tomeno, H. N. Fuke, H. Iwasaki, M. Sahashi, and Y. Tsunoda, J. Appl. Phys. \textbf{86}, 3853 (1999).
\bibitem{kaneko1987} T. Kaneko, T. Kanomata, and K. Shirakawa, J. Phys. Soc. Japan \textbf{56}, 4047 (1987); K. Takenaka, T. Inagaki, and H. Takagi, Appl. Phys. Letts. \textbf{95}, 132508 (2009);
Y. Sun, Y. Guo, Y. Tsujimoto, J. Yang, B. Shen, W. Yi, Y. Matsushita, C. Wang, X. Wang, J. Li, C. L. Satish, and K. Yamaura, Inorg. Chem. \textbf{52}, 800 (2013)
\bibitem{uchida2016} T. Uchida, Y. Kakehashi, and N. Kimura, J. Magn. Mag. Mat. \textbf{400}, 103 (2016).

%\bibitem{reimers1993} Jan N. Reimers and A.J. Berlinsky, Phys. Rev. B \textbf(48), 9539 (1993).
 %\bibitem{marty2008} K. Marty, V. Simonet, E. Ressouche, R. Ballou, P. Lejay, and P. Bordet, Phys. Rev. Lett. \textbf{101}, 247201 (2008).
%\bibitem{spinwave} M. D. Leblanc and B. W. Southern (unpublished).
%\bibitem{tsunoda2009} M. Tsunoda, H. Takahashi, and M. Takahashi, IEEE Trans. Magn. \textbf{45}, 3877 (2009).
%tusnoda2009: Systematic Study for Magnetization Dependence of Exchange Anisotropy Strength in Mn-Ir/FM (FM=Ni-Co, Co-Fe, Fe-Ni) Bilayer System
%\bibitem{tsunoda2010} M. Tsunoda et al., Appl. Phys. Lett. \textbf{97}, 072501 (2010).
%tsunoda: Linear correlation between uncompensated antiferromagnetic spins and exchange bias in Mn-Ir/Co100xFex bilayers
%\bibitem{tsunoda2012} H. Takahashi, M. Tsunoda, and M. Takahashi, IEEE Trans. Magn. \textbf{48}, 4347 (2012).
%tsunoda2012: Perpendicular Exchange Anisotropy in Mn-Ir/Fe-Co/[Pt/Co]4 Multilayers

\bibitem{MarshallandLovesey1971} W. Marshall and S. W. Lovesey, {\it Theory of Thermal  Neutron Scattering} (Clarendon Press, Oxford, 1971).
%\bibitem{FormFactor} The magnetic form factor can be computed using the ILL tables for Mn ions: http://www.ill.eu/sites/ccsl/ffacts/ffachtml.html
\bibitem{haraldsen2012} J. T. Haraldsen, R. S. Fishman, and G. Brown Phys. Rev. B, \textbf{86}, 024412 (2012).
\bibitem{brownwebsite} P. J. Brown, j0 form factors for 3d transition elements and their ions," https://www.ill.eu/sites/ccsl/ffacts/ffactnode5.html, 1998.
[Online; accessed 10-May-2016].
 


\bibitem{Granroth2006} G. E. Granroth, D. H. Vandergriff,  and S. E. Nagler,
	Physica B - Cond. Matt.\textbf{385-386},1104 (2006).

\bibitem{Granroth2010} G. E. Granroth, A. I. Kolesnikov,  T. E. Sherline, J. P. Clancy, K. A. Ross, J. P. C. Ruff, B. D. Gaulin, and  S. E. Nagler,
	J. Physics: Conf. Ser. \textbf{251}, 012058 (2010).
	
\bibitem{Lin2016} J.Y.Y. Lin, H. L. Smith, G. E. Granroth, D. L. Abernathy, M. D. Lumsden, B. Winn, A. A. Aczel, M. Aivazis, and B. Fultz,, 
Nucl. Instr. Meth.  A  \textbf{810}, 86 (2016).


\bibitem{ARNOLD2014156} O. Arnold, J.C. Bilheux, J.M. Borreguero,  A. Buts,  S.I. Campbell,  L. Chapon,  M. Doucet, N. Draper, R. Ferraz Leal, M.A. Gigg, V.E. Lynch, A. Markvardsen, 
D.J. Mikkelson, R.L. Mikkelson, R. Miller, K. Palmen, P. Parker, G. Passos, T.G. Perring, P.F. Peterson, S. Ren, M.A. Reuter, A.T. Savici, J.W. Taylor, R.J. Taylor, R. Tolchenov, W. Zhou, and J. Zikovsky,
Nucl. Instr. Meth.  A  \textbf{764}, 156 (2014).
	
\bibitem{azuah2009dave} R. T. Azuah, L. R. Kneller, Y. Qiu, P. L.W. Tregenna-Piggott, C. M. Brown, J. R. D. Copley, and R. M. Dimeo,
	J.  Res. NIST \textbf{114}, 341 (2009).
	
\bibitem{PhysRevLett.112.017205} H.Chen, Q. Niu, and A. H.  MacDonald, A. H., Phys. Rev. Lett., \textbf{112}, 
 017205 (2014).

%szunyogh2011: Atomistic spin-model based on a new spin-cluster expansion technique: Application to the IrMn3/Co interface

%\bibitem{diep1989} H. T. Diep and H. Kawamura, Phys. Rev. B \textbf{40}, 7019 (1989).
%diep: First-order phase transition in the fcc Heisenberg antiferromagnet

%\bibitem{zhitom14} M. E. Zhitomirsky, P. C. W. Holdsworth, and R. Moessner, Phys. Rev. B \textbf{89}, 140403(R) (2014).
%\bibitem{diep08} V. Thanh Ngo and H. T. Diep, Phys. Rev. E \textbf{ 78}, 031119  (2008).

\end{thebibliography}
\end{document}